\documentclass[preprint,aps]{revtex4}
\begin{document}

\title{On the Anticipatory Aspects of the Four Interactions:
what the Known Classical and Semi-Classical Solutions Teach us.}

\author{Luca Lusanna}

\affiliation
{Sezione INFN di Firenze\\ L.go E.Fermi 2 (Arcetri)\\ 50125
Firenze, Italy\\ E-mail: LUSANNA@FI.INFN.IT}

\begin{abstract}

\vskip 1truecm

The four (electro-magnetic, weak, strong and gravitational)
interactions are described by singular Lagrangians and by
Dirac-Bergmann theory of Hamiltonian constraints. As a consequence
a subset of the original configuration variables are {\it gauge
variables}, not determined by the equations of motion. Only at the
Hamiltonian level it is possible to separate the gauge variables
from the deterministic physical degrees of freedom, the {\it Dirac
observables}, and to formulate a well posed Cauchy problem for
them both in special and general relativity. Then the requirement
of {\it causality} dictates the choice of {\it retarded} solutions
at the classical level. However both the problems of the classical
theory of the electron, leading to the choice of ${1\over 2}\,
(retarded + advanced)$ solutions, and the regularization of
quantum field teory, leading to the Feynman propagator, introduce
{\it anticipatory} aspects. The determination of the relativistic
Darwin potential as a semi-classical approximation to the
Lienard-Wiechert solution for particles with Grassmann-valued
electric charges, regularizing the Coulomb self-energies, shows
that these anticipatory effects live beyond the semi-classical
approximation (tree level) under the form of radiative
corrections, at least for the electro-magnetic interaction.
\bigskip

Talk and "best contribution" at The Sixth International Conference
on Computing Anticipatory Systems CASYS'03, Liege August 11-16,
2003

\end{abstract}

\maketitle

\section{Introduction}

Our understanding of the four interactions (gravitational,
electro-magnetic, weak and strong, with or without super-symmetry)
has led to a description of physics which utilizes classical
action principles whose associated Lagrangian densities are
singular. Therefore the equations of motion for the fields and/or
the particles cannot be written in {\it normal} form, namely they
cannot be solved for the accelerations and put in the form of the
equations of Newtonian mechanics ($m \vec a = \vec F$). This is
due to the requirements of manifest Lorentz covariance and gauge
invariance for special relativistic systems and of general
covariance for general relativistic systems. To comply with them
we have to introduce redundant non-physical quantities which
transform covariantly like tensors under the appropriate group. In
special relativistic particle theory we must use gauge potentials
$A_{\mu}(x)$ defined modulo gauge transformations and particle
world-line coordinates (events) $x^{\mu}_i(\tau )$, $i=1,..,N$,
defined modulo reparametrizations of the scalar affine parameter
$\tau$. While in the first case two components of the gauge
potential are non-measurable mathematical quantities, in the
second case we have as many time coordinates $x^o_i(\tau )$ as
particles and we must connect them to the {\it time} of the
special relativistic clocks in such a way that physical quantities
(for instance bound states of the particles) are independent from
the unphysical {\it relative times} $x^o_i(\tau )-x^o_j(\tau )$.
The description of bound states must be independent from the
freedom of the observer of looking at the particles at the same
time or with any prescribed time delay among them \cite{1}. With
strong and electro-weak interactions we must use the non-Abelian
Lie algebra-valued gauge potentials $A_{a\mu}(x)\, T^a$, with
$T^a$ a matrix representation of either $SU(3)$ or $SU(2) \times
U(1)$, and again for each value of $a$ two components of the gauge
potential are gauge variables \cite{2}.

Einstein's general relativity and all the existing variants
emphasize that, contrary to special relativity where Minkowski
space-time is absolute, space-time points loose their physical
individuality (the {\it Hole Argument}\cite{3}), that coordinates
are purely conventional and that measurable physical quantities
must not depend on their choice. These are consequences of the
invariance of the theory under active diffeomorphisms and general
coordinate transformations (passive diffeomorphisms) and this
implies the general covariance of the equations of motion (they
must take the same functional form in every coordinate system).
This leads to the fact that only two of the ten components of the
metric tensor are dynamically determined by Einstein's equations
in every coordinate system. As a consequence there is a going on
debate on which are the {\it observables} in general relativity
(see for instance Ref.\cite{4}), in particular for the
gravitational field itself, and on how to build a reference
standard (Global Positioning System?) for an empirical
determination of a system of coordinates with the associated
metric tensor to be used for every subsequent measurement of
times, lengths, angles, tidal and inertial effects of the
gravitational field, matter properties. In Refs.\cite{5} there is
a complete discussion of these problems and a proposal for their
solution based on a formulation of metric \cite{6} and tetrad
\cite{7,8} gravity in globally hyperbolic, topologically trivial,
asymptotically flat at spatial infinity space-times. It is shown
that in these space-times, belonging to the family of
Christodoulou - Klainermann space-times \cite{9}, it is possible
to define a rest-frame instant form of gravity, where the
evolution is governed by the ADM energy (they are a
counter-example to the {\it frozen picture without evolution} of
general relativity).

\section{Gauge Variables and Dirac Observables}

The main consequence of this state of affairs is that both in
special and general relativity the {\it covariant} equations of
motion do not determine the unphysical gauge degrees of freedom
present in the covariant description of the system. This fact
gives rise to a {\it lack of determinism} in the time evolution of
the covariant variables. As a consequence, one has to face the
problem to disentangle the arbitrary gauge variables from the
deterministic physical measurable quantities ({\it Dirac
observables}), whose equations of motion in the form of a
hyperbolic system of (partial) differential equations with a well
posed Cauchy problem on a space-like initial hyper-surface of
space-time. Once this is done, the next problem is the empirical
impossibility of knowing (i.e. preparing) all the physical initial
data on a non-compact Cauchy surface: this the price to be paid to
get the theorems on the existence and uniqueness of the solutions
of the equations of motion.

The natural environment where to formulate, study and try to solve
these problems at the relativistic level is not configuration
space but phase space. Here the theory of singular Lagrangians,
due to the second Noether theorem\cite{10}, gives rise to
Dirac-Bergmann theory of Hamiltonian constraints \cite{11} by
means of which a well defined formulation of the lack of
determinism of every special or general relativistic theory can be
given in the framework of Hamilton-Dirac equations of motion.
Moreover the family of Shanmugadhasan canonical transformations
\cite{12} allows, at least at a heuristic level in field theory,
to find special canonical bases in which: a) a subset of the new
momenta are the generators of the Hamiltonian gauge
transformations (Abelianized form of the first class constraints);
b) the conjugated configurational variables are an Abelianized
form of the redundant gauge variables; c) the remaining pairs of
canonical variables form a Darboux basis of {\it Dirac
observables} (no method is known in configuration space to find a
basis of observables). The Dirac observables are the physical
measurable quantities which have a deterministic evolution, since
they satisfy hyperbolic Hamilton equations with a completely
determined Hamiltonian and with a well posed Cauchy problem.
However, in general, they are not covariant, namely their
functional form is gauge dependent. In special relativity it is
known how to find covariant functions of the Dirac observables
(for instance in electromagnetism such functions are the
transverse electric and magnetic fields). Instead in general
relativity it is still an open problem how to build functions of
the Dirac observables of the gravitational field which do not
depend on the choice of the coordinates. In the first of
Refs.\cite{5} there is the conjecture that in gravity there are
special Shanmugadhasan canonical bases, in which both the Dirac
observables and the gauge variables (describing generalized tidal
and inertial effects respectively) are coordinate-independent:
they should emerge from a Hamiltonian re-formulation of the
Newman-Penrose approach \cite{13}.

\section{The Initial Value Problem and Anticipation}

Modulo manifest covariance Dirac observables are the deterministic
relativistic counterpart of the configurational or phase space
variables of ordinary Newton mechanics, both classically and
quantum mechanically. Therefore, it is only at the level of Dirac
observables that the problematic about {\it anticipation} is well
posed (see the review paper of Dubois\cite{14} on anticipatory
systems). Moreover, numerical simulations and/or approximations
with finite difference equations, with their possible change of
perspective on the properties of solutions, are physically
motivated only for the Hamilton equation for Dirac observables.

Given the deterministic equations of motion of an isolated system
we have first of all to choose definite {\it boundary conditions}
so that the functional space in which to look for solutions is
well defined (this step usually is trivial in mechanical
problems). Then we have to formulate a well posed {\it initial
value problem}. The two main classes of such problems are
connected with the names of Cauchy and Dirichlet. In a Cauchy
problem for a second order (partial) differential equation we have
to give the initial configuration and the initial velocities on
the space-like hyper-surface corresponding to the initial time (in
relativistic theories this implies a choice of {\it equal time}
(simultaneity) Cauchy surface and the description of the
relativistic system is acceptable if the physical results do not
depend on this choice). Then {\it causality} induces the choice of
{\it retarded} Green functions in the construction of solutions to
the given problem, which therefore do not depend on events in the
absolute future of the initial configuration of the system.
However, as noted in Ref.\cite{14}, for simple mechanical systems
the equations of motion may depend on parameters, which a
posteriori can be seen to coincide with the value of some
configurational variable at a later final time. In these cases,
some type of anticipation is present in the system.

Instead in a Dirichlet problem we look for solutions of the
equations of motion determined by an initial configuration of the
system at time $t_i$ and by a final one at time $t_f > t_i$.
Clearly in this type of solutions we have {\it anticipation},
since the knowledge of the future configuration determines the
intermediate configurations. The formulation of Dirichlet problems
is a much harder task than the Cauchy problem, since all the
topological properties of the intermediate configurations have to
be known in advance.

Classical relativistic theories, relying on the standard notion of
causality, tried to avoid anticipation by using only retarded
solutions of the classical relativistic wave equations (in them
there are no parameters hiding anticipation). However this attempt
was frustrated by the problems of the classical theory of the
electron. The Abraham-Lorentz-Dirac equation, giving the correct
Larmor formula for the radiation in wave zone, depends on the time
derivative of the acceleration. To avoid {\it runaway solutions},
with their unlimited growing of the acceleration, we have to
re-formulate the equation as an integral equation, and this
creates a problem with causality (the {\it pre-acceleration}, see
for instance Ref.\cite{15}), which is usually solved by refuting
the validity of the classical theory and invoking quantum
mechanics. Moreover, in the limit of a point-like electron we have
the appearance of the infinities connected with the Coulomb
self-energies.

These problems of the classical theory were debated for a century
and the only way out was Feynman-Wheeler electrodynamics \cite{16}
with their theory of the absorbers. Here {\it anticipation is
present}, because the chosen solutions admit a symmetric
combination of {\it retarded} and {\it advanced} effects.

The whole problematic was re-formulated with quantum field theory
regularized in such a way to eliminate the infinities. This is a
theory {\it with anticipation}, because, due to the fact that no
one has been yet able to regularize the product of two retarded
distributions, it uses Feynman Green functions, whose real part is
the symmetric combination of retarded and advanced Green functions
(the imaginary part is connected to diffractive absorption in the
other channels of the many-body theory). The same happens in the
formulation of the theory by means of path integrals, since they
correspond to a Dirichlet problem, in which the initial and final
configurations are given. Let us remark that, notwithstanding all
the successes of quantum field theory, we do not yet know whether
the causal problems of the classical theory are solved or not. In
perturbative quantum electrodynamics they are absent by
construction. But the perturbative expansion is only an asymptotic
series which cannot be resummed. On the other hand at the
non-perturbative level (the path integral approach) we do not know
how to formulate the problem.

Moreover quantum field theory is a special relativistic theory and
it is not known how to incorporate gravity in it in a regularized
way which takes into account the already quoted problems of the
physical identification of space-time points. See the first of
Refs.\cite{5} for a proposal of quantization of gravity preserving
relativistic causality.

Indeed our understanding of quantum theory requires that the
events (the points) of space-time and the simultaneity Cauchy
surface be fixed in advance in an absolute way. For instance there
are suggestions (see Penrose contribution in Ref.\cite{17}) that
gravity may play a role in understanding the unsolved problem of
how the {\it potentialities} of quantum theory (which gives only
probabilities) become the {\it actualities} we experience (the
problem of the collapse of the wave function when we do a
measurement on an entangled state, consequence of the linearity of
the Scrhoedinger equation and of the superposition principle).
Another unsolved problem is the integrability of 4-dimensional
classical and quantum field theories and the repercussions of
chaotic motions on both the regularizzability of quantum field
theories and on our understanding of gravitational physics.

Therefore our present knowledge about physical systems points in
the direction of the presence of {\it anticipation}, but also
stimulates us to understand better the interface between classical
and quantum systems, especially at the relativistic level in view
of the inclusion of gravity.

\section{The Rest-Frame Instant Form of Dynamcs}

Therefore I started a research program aiming to arrive at a
unified description of the four interactions at the classical
level in terms only of Dirac observables (see Ref.\cite{2} for a
review). The first stage was to re-formulate all classical
isolated systems (particles, strings, fields) on arbitrary
space-like hyper-surfaces (corresponding to a concept of
simultaneity for a family of time-like non-inertial accelerated
observers) instead that on space-like hyper-planes (inertial
observers). This was done in such a way that physical results are
independent from the choice of the hyper-surface (special
relativistic general covariance of parametrized Minkowski
theories). In this way it is possible to add the gravitational
field to the description in such a way that the switching off of
Newton constant allows to recover the isolated system without
gravity (solution of the deparametrization problem of general
relativity).

Then I have studied the description of the isolated systems in
special relativity when they are restricted to the special family
of space-like hyper-planes orthogonal to the conserved 4-momentum
of the system. These so-called {\it Wigner hyper-planes} are
determined by the isolated system itself and define its intrinsic
rest frame \cite{18}. They opened a new viewpoint to understand
the old unsolved problem of which is the best definition of the
relativistic center of mass and allowed the identification of new
kinematical variables both for the N-body problem and for extended
systems \cite{19}. Moreover in this way it was possible to define
a new instant form of dynamics in Dirac sense \cite{20}, the {\it
Wigner covariant 1-time rest-frame instant form} \cite{18,2} and
to identify an {\it intrinsic classical unit of length} for
extended rotating systems, the M\o ller radius \cite{21}, which
hopefully will be used as a {\it physical ultraviolet cutoff} to
regularize infinities \cite{2}. At the classical level the M\o
ller radius identifies a world-tube which contains the region of
{\it non-covariance} of the relativistic center of mass (the
classical counterpart of the Newton-Wigner position operator) and
it can be shown to be the remnant in flat space-time of the energy
conditions of general relativity. At the quantum level it is
proportional to the Compton wavelength of the system, so that it
impossible to localize the non-covariant center of mass without
producing quantum pairs. Therefore, the world-tube, which is a
consequence of the existence of the light cone (Lorentz signature
of space-time, constancy of the velocity of light), lies at the
intersection between quantum theory and general relativity.

The next step of the program was the determination of the Dirac
observables of the various theories \cite{2}. This has been
accomplished for the standard  SU(3)xSU(2)xU(1) model of
elementary particles (electro-magnetic, weak and strong
interactions) in special relativity on the Wigner hyper-planes.
Now, after the definition of the rest-frame instant form of metric
\cite{6} and tetrad \cite {7,8} gravity, the Dirac observables for
the gravitational field are under investigation and we begin to
understand how to identify the points of space-time a posteriori
by using the gravitational field. A first byproduct has been the
definition of {\it background-independent gravitational waves} in
post-Minkowskian space-times \cite{22} by means of a Hamiltonian
linearization of tetrad gravity in a completely fixed non-harmonic
3-orthogonal gauge. The final step will be to put together the
standard model of elementary particles and the gravitational
field. Then one will face the problem of quantization.

\section{A Semi-Classical Description of Charged Particles plus
the Electromagnetic Field}

In the meanwhile I revisited the classical problem of charged
particles plus the electro-magnetic field, which was the source of
anticipation in physics, in the rest-frame instant form on Wigner
hyperplanes \cite{23,24}. To avoid infinities in the classical
self-energies, in place of an extended electron model a {\it
semi-classical} description of the electric charge with Grassmann
variables was used \cite{25}. Experimentally the electric charge
is quantized: if magnetic monopoles exist the product of the
electric and magnetic charges is proportional to the Planck
constant; if they do not exist, we do not understand the quantum
of electric charge. If we consider a quantum operator (like the
spin or the electric charge) whose spectrum has a finite number of
discrete levels, we cannot recover the classical theory by going
to high quantum numbers. However, we can get a semi-classical
description by replacing the operator with a suitable Grassmann
variable (Grassmann variables $Q_i$ are mathematical objects
satisfying $Q_iQ_j+Q_jQ_i=0$ so that $Q_i^2=0$): in this way we
get a consistent treatment in which the quantum operator $\hat Q$
is replaced by an infinitesimal quantity $Q$ (an infinitesimal
quantum) and infinitesimals of higher order are discarded
($Q^2=0$)\cite{26}. With an appropriate quantization rule we can
recover the quantum theory with the operator $\hat Q$. On the
other hand, by means of Berezin-Marinov Grassmann-valued density
matrix \cite{27} we can get the classical theory: for the spin of
a particle there is no classical analogue, while it exists for the
electric charge \cite{25}.

Therefore we studied the system of N positive energy particles
with Grassmann-valued electric charges $Q_i$ ( $Q^2_i=0$,
$Q_iQ_j=Q_jQ_i\not 0$ for $i\not= j$) coupled to a dynamical (not
external) electro-magnetic field, both in the case of spinless
particles \cite{23} and of spinning ones \cite{24} with
Grassmann-valued spins \cite{28}, in the rest-frame instant form.
By means of a Shanmugadhasan canonical transformation the system
was expressed only in terms of Dirac observables both for the
particles (${\vec \eta}_i(\tau )$, ${\vec \kappa}_i$ with
$x^{\mu}_i(\tau )=z^{\mu}(\tau ,{\vec \eta}_i(\tau ))$, where the
functions $z^{\mu}(\tau ,\vec \sigma )$ describe the embedding of
the Wigner hyper-planes in Minkowski space-time) and for the
electro-magnetic field (it corresponds to the {\it radiation
gauge} with transverse vector potential ${\vec A}_{\perp}(\tau
,\vec \sigma )$ and electric fields ${\vec E}_{\perp}(\tau ,\vec
\sigma )$).

A first consequence of the Grassmann-valued charges ($Q^2_i=0$,
$Q_iQ_j=Q_jQ_i\not= 0$) is the regularization of the Coulomb
self-energies (the $i\not= j$ rule) in the rest-frame Hamiltonian.
This Hamiltonian is the rest-frame invariant mass of the isolated
system. In the case of positive energy spinning particles it was
necessary to make a semi-classical Foldy-Wouthuysen transformation
to determine how the spin of the positive energy particles couples
to the electric field. The form of the Hamiltonian in the two
cases is [${\vec \xi}_i$ are the Grassmann variables for the
description of the semi-classical spin $S^r_i=-{i\over 2}
\epsilon^{ruv} \xi_i^u\xi_i^v$]

\begin{eqnarray*}
H_{spinless} &=&\sum_{i=1}^{N}\sqrt{m_{i}^{2}+(\check{\vec{%
\kappa}}_{i}(\tau )-Q_{i}{\check{\vec{A}}}_{\perp }(\tau ,\vec{\eta}%
_{i}(\tau )))^{2}}+  \nonumber \\
&+&\sum_{i\neq j}\frac{Q_{i}Q_{j}}{4\pi \mid \vec{\eta}_{i}(\tau )-\vec{\eta}%
_{j}(\tau )\mid }+\int d^{3}\sigma
{\frac{1}{2}}[\check{\vec{E}}_{\perp
}^{2}+\check{\vec{B}}^{2}](\tau ,\vec{\sigma}),  \nonumber \\
 &&{} \nonumber \\
 H_{spin}  &=&\sum_{i=1}^{N}\Big[\sqrt{m_{i}^{2}+(\check{%
\vec{\kappa}}_{i}(\tau )-Q_{i}{\check{\vec{A}}}_{\perp }(\tau ,\vec{\eta}%
_{i}(\tau )))^{2}}-  \nonumber \\
&+&i\frac{Q_{i}\vec{\xi}_{i}(\tau )\times \vec{\xi}_{i}(\tau )\cdot \check{%
\vec{B}}(\tau ,\vec{\eta}_{i}(\tau ))}{2\sqrt{m_{i}^{2}+\check{\vec{\kappa}}%
_{i}^{2}(\tau )}}-i\frac{Q_{i}\check{\vec{\kappa}}_{i}(\tau )\cdot \vec{\xi}%
_{i}(\tau )\,\vec{\xi}_{i}(\tau )\cdot \check{\vec{E}}_{\perp }(\tau ,\vec{%
\eta}_{i}(\tau ))}{(m_{i}+\sqrt{m_{i}^{2}+\check{\vec{\kappa}}_{i}^{2}(\tau )%
})\sqrt{m_{i}^{2}+\check{\vec{\kappa}}_{i}^{2}(\tau )}}\Big]+
\nonumber \\
&+&\sum_{i\neq j}\Big[\frac{Q_{i}Q_{j}}{4\pi \mid \vec{\eta}_{i}(\tau )-\vec{%
\eta}_{j}(\tau )\mid }-  \nonumber \\
&-&i\frac{Q_{i}Q_{j}\check{\vec{\kappa}}_{i}(\tau )\cdot
\vec{\xi}_{i}(\tau
)\,\vec{\xi}_{i}(\tau )\cdot (\vec{\eta}_{i}(\tau )-\vec{\eta}_{j}(\tau ))}{%
4\pi \mid \vec{\eta}_{i}(\tau )-\vec{\eta}_{j}(\tau )\mid ^{3}(m_{i}+\sqrt{%
m_{i}^{2}+\check{\vec{\kappa}}_{i}^{2}(\tau )})\sqrt{m_{i}^{2}+\check{\vec{%
\kappa}}_{i}^{2}(\tau )}}\Big]+  \nonumber \\
&+&\int d^{3}\sigma {\frac{1}{2}}[\check{\vec{E}}_{\perp }^{2}+\check{\vec{%
B}}^{2}](\tau ,\vec{\sigma}).
\end{eqnarray*}

We then studied the coupled Hamilton equations for the particles
and the electro-magnetic field. By integrating the equations for
the electro-magnetic field, we can express the transverse vector
potential as the sum of a pure radiation term plus the transverse
Lienard-Wiechert term \cite{29}, which depends on the particles
and on the choice of the Green function (retarded, advanced,
symmetric,..). Due to the semi-classical approximation $Q^2_i=0$,
each particle creates a transverse Grassmann-valued vector
potential, so that a single particle cannot irradiate energy (it
is a quantity quadratic in the fields) since it is of order
$Q^2_i=0$. This solves at the semi-classical level the causality
problems of the Abraham-Lorentz-Dirac equation. However, if we
have various particles and we do not use their equations of motion
(namely we consider them as sources), then there is emission of
energy (it is of order $Q_iQ_j$, coming from interference terms)
and we recover the Larmor formula in the wave zone. This is enough
to explain the asymptotic radiation from a region containing
macroscopic sources.

Independently from that, we can consider the Lienard-Wiechert
solutions at some point P of space-time: they depend on the
delayed (retarded, advanced, symmetric,..) times of the particles
creating them. We can develop these delayed times so to re-express
the Lienard-Wiechert solutions only in terms of quantities
evaluated on the Wigner hyperplane through the point P. A priori
we get expressions which depend on the transverse potential at
that time and on its time derivatives of any order (ordinary and
higher accelerations) evaluated at that time (like it happens in
Feynman-Wheeler electrodynamics if one make an equal time
development of its integro-differential equations of motion).
However, by using the equations of motion of the particles, we can
show that only the transverse vector potential and its first time
derivative (the velocity) survive: all the accelerations are
multiplied by $Q^2_i=0$. Moreover, we get the same result
whichever is the Green function considered (retarded, advanced,
symmetric,..). It is now possible to express the semi-classical
Lienard-Wiechert transverse vector potential and the electric and
magnetic fields in terms of the positions and momenta of the
particles. Therefore, Grassmann-valued electric charges allow to
extract the {\it semi-classical action-at-a-distance potential}
hidden in the electro-magnetic interaction notwithstanding it is
an interaction with delay, so that {\it no anticipation} survives
at the semi-classical level.

Then we make a canonical reduction to the sector of bound states,
in which only particle degrees of freedom are present. This is
done by forcing the electro-magnetic fields in the Hamiltonian to
coincide with the semi-classical Lienard-Wiechert solution. After
having found the new canonical variables for the particles in this
reduced phase space, we can re-express the Hamiltonian in terms of
them. The final reduced Hamiltonian contains: a) the relativistic
kinetic energy of the positive energy particles; b) the
regularized Coulomb potential; c) the semi-classical Darwin
potential with all its relativistic corrections. While in the
spinless case the lowest order in $1/c^2$ of the semi-classical
Darwin potential coincides with the standard Darwin potential
\cite{30}, in the spinning case at the same order we get an
expression for such potential, which in the 2-body case with
arbitrary masses after quantization coincides with the potential
which was obtained from Bethe and Salpeter \cite{31} starting from
quantum field theory in an instantaneous approximation. All the
spin-orbit and spin-spin terms for positronium ($m_1=m_2$),
muonium ($m_1\not= m_2$) and hydrogen-like atoms ($m_1\rightarrow
\infty$) are reproduced.

\section{Conclusion}

Therefore we have discovered a way to find the semi-classical
potential for a Cauchy problem which corresponds to the static and
non-static contributions coming from the Feynman diagrams with an
one-photon exchange. Let us remark that Feynman diagrams
correspond to a Dirichlet problem which has {\it anticipation}: we
are able to re-formulate it as a Cauchy problem {\it without
anticipation} at the semi-classical level. Anticipation is pushed
to the radiative corrections and many-photon exchanges, which take
into account the delay.

We are now planning to redo the calculations on arbitrary
space-like hyper-sufaces to study how accelerated observers
describe the radiation emitted by particles. We will also try to
extend the method to the quark model (strong interaction bound
states of quarks with Grassmann-valued color charges \cite{32}) to
see whether it is possible to demonstrate the confinement of
quarks with these semi-classical methods. The analogous problem in
linearized tetrad gravity \cite{22} with a perfect fluid as matter
will push us to find the physical Hamiltonian of our gauge
containing the relativistic action-at-a-distance potentials and
the tidal interactions, and to determine the relativistic
quadrupole emission formula. This would clarify the status of
anticipatory effects in gravitational physics.

In conclusion we begin to understand some unexpected features of
the semi-classical approximation in the rest-frame instant form
like the elimination of anticipation. The role played by the
Grassmann quantities is intriguing, because we do not understand
the quantization of charges and which is the connection of their
regularizing properties with the standard procedure of
regularization in quantum field theory. Maybe the presence of a
physical cutoff (the M\o ller radius) will help to clarify the
last point. Is it possible to go beyond the semi-classical
approximation and to replace the anticipatory many-photon
exchanges and the radiative corrections with higher order
potentials for a Cauchy problem without anticipation? Is a
consistent addition of gravity going to change things
qualitatively? Does the gravitational mass (the charge of
gravitation) arise through some quantization and symmetry breaking
mechanisms as it is suggested by special relativistic particle
theory or has a completely different origin and a completely
different semi-classical behaviour? No answer to these questions
is known at present.

\end{document}